\documentstyle[12pt]{article}
\begin{document}

\title{COHERENT CURRENT STATES IN MESOSCOPIC FOUR-TERMINAL
JOSEPHSON JUNCTION}
\author{ Malek Zareyan~${}^{a}$ and A.N.\ Omelyanchouk~${}^{b}$
\\
{\small {\em ${}^{a}$ Institute for Advanced Studies in Basic Sciences,}}
\\
{\small {\em 45195-159, Gava Zang, Zanjan, Iran}}
\\
{\small {\em ${}^{b}$ B.Verkin Institute for Low Temperature Physics and
Engineering,}}
\\
{\small {\em National Academy of Sciences of Ukraine,}}
\\
{\small {\em 47 Lenin Ave., 310164 Kharkov, Ukraine}}}
\maketitle
\date{}
\begin{abstract}

A theory is offered for the ballistic 4-terminal Josephson junction.
The studied system consists of a mesoscopic two-dimensional normal
rectangular layer which is attached in each side to the bulk
superconducting banks (terminals). The relation between the currents
through the different terminals, which is valid for arbitrary temperatures 
and junction sizes, is obtained. The nonlocal coupling of the supercurrents 
leads to a new effect, specific for the mesoscopic weak link between two 
superconducting rings; an applied magnetic flux through one of the rings 
produces a magnetic flux
in the other ring even in the absence of an external flux through the other
one.
The phase dependent distributions of the local density of Andreev states,
of the supercurrents and of the induced order parameter are obtained. The
"interference pattern" for the anomalous average inside the two dimensional
region can be regulated by the applied magnetic fluxes or the transport 
currents.
For some values of the phase differences between the terminals, the current
vortex state and the two dimensional phase slip center are appeared.
\end{abstract}

\newpage

\section{Introduction}
The Josephson multiterminal junction presents the microstructure in which
the weak coupling takes place between several massive superconducting banks
(terminals)[1-3]. Compared with the conventional (2-terminal)
Josephson junctions \cite{Bar} such systems have additional degrees of
freedom and corresponding set of the control parameters. With the result, for
example, the current- or voltage- biased and the magnetic flux-driven
regimes can be combined in one multiterminal junction. The specific
multichannel interference effects were studied theoretically and
experimentally in the novel superconducting device, the 4-terminal SQUID
controlled by the transport current [5-7].
Recently another system based on a Josephson 4-terminal junction was
studied \cite{OOV98}. It consists of two superconducting rings, each
interrupted by a Josephson junction, which are at the same time weakly
coupled with each other. The macroscopic quantum states of such composite
system can be regulated by the difference of the magnetic fluxes applied
through the rings, in analogy with the phase difference between two weakly
coupled bulk superconductors. The nonlinear coupling via the Josephson 
4-terminal leads to the cooperative behaviour of the rings in some region of
the applied magnetic fluxes, which was called \cite{OOV98} magnetic 
flux locking.

The 4-terminal junction, which was studied in Refs. [5-8], is a
system of short microbridges going from a weak point to massive
superconducting banks (Fig.1a). The order parameter (both its amplitude and
phase) in the common centre is a function of the currents through all the
microbridges. The supercurrent flowing into the $i$th bank is determined by
the phases of the order parameter $\varphi _i,(i=1,..4)$ in 
all the banks \cite{OOV95}:

\begin{equation}
I_i=\frac{\pi \Delta _0^2(T)}{4eT_c}\frac 1{\sum_j1/R_j}\sum\limits_j\frac
1{R_iR_j}\sin (\varphi _i-\varphi _j). 
\end{equation}
The relation (1) was obtained in the frame of the Ginzburg-Landau approach,
which is valid for temperatures $T$ close to the critical temperature $T_c$. 
As it was pointed in the Ref. \cite{OOV98}, the macroscopic interference effects due to
coupling of supercurrents in different terminals are not
restricted by the special kind of the 4-terminal junction (Fig.1a). In fact,
any mesoscopic 4-terminal weak link will produce a coupling similar to the
relation (1). In the present paper, the microscopic theory of the mesoscopic 
ballistic 4-terminal junction is developed. We consider a Josephson weak
coupling through the two-dimensional normal layer which is connected with
four bulk superconducting terminals as it is shown in Fig.1b. Such a S-2DEG-S
structure was experimentally realized in Ref. \cite{Kla} for the case of two
terminals. It was shown in \cite{Kla} that this new class of fully phase 
coherent Josephson junctions demonstrate the nonlocal phase dependence of
mesoscopic supercurrents. We study the coherent current states in such a
4-terminal structure within the quasiclassical equations for transport-like
Green's functions. The relation between the currents in the different
terminals,
which is valid for arbitrary temperatures and junction sizes, is obtained.
The structure of current carrying states inside the mesoscopic 4-terminal
junction presents the interest itself. 
As it is well known (see e.g. \cite{Bag1}), in
ballistic Josephson junction with direct conductivity the supercurrent
flows through the local Andreev levels. In multiterminal case
considered here, the spatial distribution of current density and of 
the order parameter, and hence the phase-dependent Andreev levels, are 
determined by the phase differences between all terminals. Thus, they 
can be regulated by
the external control parameters, i.e. the transport currents and (or)
the applied
magnetic fluxes. In Section 2, we present the description of the system and
formulate basic equations and boundary conditions. In Section 3 the
current-phase relations analogous to (1) are derived for the cases of small
(as compared to the coherence length) and also arbitrary 
junction sizes. The spatial
distributions of the supercurrent density and of the induced order
parameter are studied in Section 4.

\section{Model and basic equations.}

The studied system consists of 4 bulk superconducting banks which are
contacted with 4 sides of rectangular two-dimensional (2D) normal layer
having the length $L$ and width $W$ (see Figs. 1b and 2). The sizes $L$ 
 and $W$ are supposed to be much larger than Fermi 
wave-length $\lambda _F=h/p_F$ . To study the stationary
coherent current states in 4-terminal ballistic
junction we use the Eilenberger equations \cite{Eil} for $\xi $-integrated
Green's functions: 
\begin{equation}
{\bf v}_F\frac \partial {\partial {\bf r}}\hat G+[\omega \hat \tau _3+\hat \Delta , \hat G]=0, 
\end{equation}
where

$$
\hat G_{\omega}({\bf v}_F{\bf ,r})=\left( 
\begin{array}{cc}
g_\omega & f_\omega \\ 
f_\omega ^{+} & -g_\omega 
\end{array}
\right) 
$$
is the matrix Green's function, which depends on the Matsubara frequency $%
\omega $, the electron velocity on the Fermi surface ${\bf v}_F$ and the
coordinate ${\bf r}$;

$$
\hat \Delta =\left( 
\begin{array}{cc}
0 & \Delta  \\ 
\Delta ^{*} & 0
\end{array}
\right)  
$$
is the superconducting pair potential. For the self-consistent off-diagonal
potential $\Delta ({\bf r)\ }$and current density ${\bf j(r)\ }$ we have the
expressions

\begin{equation}
\Delta({\bf r}) =\lambda 2\pi T\sum\limits_{\omega >0}\left\langle f_\omega
\right\rangle ,
\end{equation}

\begin{equation}
{\bf j}({\bf r})=-4\pi ieN(0)T\sum\limits_{\omega >0}\left\langle {\bf v}_Fg_\omega
\right\rangle .
\end{equation}
They determine the induced order parameter $\Psi $$\equiv $$\Delta /\lambda $
and the 2D current density in the normal layer;  $N(0)=\frac m{2\pi },$ $%
\left\langle ...\right\rangle $ is the averaging over directions of 2D
vector ${\bf v}_F$, $\lambda $ is the constant of electron-phonon coupling. 

The equations (2) are supplemented by the values of $\Delta $ and Green's
functions in bulk banks far from the S-N interfaces:

\begin{equation}
\hat \Delta _i=\Delta _0(\hat \tau _1\cos \varphi _i-
\hat \tau _2\sin \varphi _i), \ \ \  \hat G_i=\frac{%
\omega \hat \tau _3+\hat \Delta _i}{\sqrt{\omega ^2+\Delta _0^2}},\ \ \
 i=1,..4.
\end{equation}
We solve Eqs. (2) by integrating over the ''transit'' trajectories of the
ballistic flight of electrons from one bank to another \cite{KO}. These
trajectories (characteristics of the differential Eqs.(2) ) are straight
lines along the direction of electron velocity (see Fig. 2). In the bulk
superconducting banks the order parameter can be taken as the constant value
(5) up to the S-N interface. In contrast to the case of 2D banks these
''rigid'' conditions for $\Delta$ \cite{Lik,KO} are valid for arbitrary sizes 
$L$ and $W$ compared with the coherence length $\xi _0\sim v_F/\Delta _0$, 
and not only for $L,W\gg\xi _0$. At the same time, the Green's function 
along the given transit trajectory varies in a distance of about $\xi _0$
where approaching the S-N interface. \par
Let us introduce the time of flight along the trajectory,
${\bf v}_F\frac \partial {\partial {\bf r}}\equiv \frac{d}{dt}$, \\ 
$t_i< t<\infty$ where $t=t_i$ corresponds to the point on $i$th S-N 
boundary and $t=\infty$ to the point inside $i$th bank far from the S-N
boundary. Then the general solution of Eqs.(2) inside the $i$th bank 
satisfying the boundary conditions (5) will be

$$
\hat G_i(t)=\frac{\omega \hat \tau _3+\hat \Delta _i}\Omega 
$$
\begin{equation}
+C_i[\Delta _0\hat \tau _3-(\omega
\cos \varphi _i+isign({\bf v}_F{\bf n}_i)\Omega \sin \varphi _i) \hat \tau
_1
\end{equation}
$$
+(\omega \sin \varphi _i-isign({\bf v}_F{\bf n}_i)\Omega \cos \varphi
_i) \hat \tau _2]e^{-2\Omega (t-t_i)}.
$$
Here ${\bf n}_i$ is the outer normal
to the $i$th side of the rectangular boundary and $\Omega =\sqrt{\omega
^2+\Delta _0^2}$.  The arbitrary constants $C_{i}$ must be found by
matching of Green's functions at in-coming and out-going points at S-N
boundaries with the solution inside the normal layer along the trajectory 
which connects these points (see Fig. 2). We consider here the simplest 
case when only Andreev reflection \cite{And} occurs at S-N interface. In
more realistic case, when usual reflection (e.g. due to the potential 
barrier) or interface roughness are present, more general matching 
conditions must be used (see \cite{OOM}).

\section{Current-Phase relations}

Inside the normal layer ($\Delta=0$), the Eilenberger equations can be 
solved analytically. If we classify the electronic trajectories inside 
the normal layer according to the sides at which they come in and go out, 
then the solution of eq. (2) can be written as
$$
\hat G_{i\rightarrow j}(t)=\frac{\omega}{\Omega}\hat \tau_3+\frac{\Delta_0}
{\Omega}\{\cosh{[2\omega (t-t_i)-i\varphi_i]} \hat \tau_1-i\sinh{[2\omega 
(t-t_i)-i\varphi_i]} \hat \tau_2\}
$$
$$
+A_{i\rightarrow j}\{\Delta_0 \hat \tau_3- [\omega\cosh{(2\omega (t-t_i)
-i\varphi_i)} +\Omega\sinh{(2\omega (t-t_i)-i\varphi_i)}]\hat \tau_1
$$
\begin{equation}
+i[\Omega\cosh{(2\omega (t-t_i)-i\varphi_i)}+\omega\sinh{(2\omega (t-t_i)
-i\varphi_i)} ]\hat \tau_2\},
\end{equation}
where $\hat G_{i\rightarrow j}(t)$ is the matrix Green's function 
along the trajectory originating in the $i$th side and extending to the
$j$th side (see Fig. 2). We denote this trajectory by $i\rightarrow j$. 
Matching (7) with solution in the banks 
(6), the corresponding $A_{i\rightarrow j}$ is obtained    
\begin{equation}
A_{i\rightarrow j}=\frac{\frac{\Delta_0}{\Omega}\sinh{(\omega t_{ji}
+i\frac{\varphi_{ji}}{2})}}{\omega\sinh{(\omega t_{ji}
+i\frac{\varphi_{ji}}{2})} +\Omega\cosh{(\omega t_{ji}
+i\frac{\varphi_{ji}}{2})}},
\end{equation}
where $t_{ji}=t_j-t_i$ and $\varphi_{ji}=\varphi_{j}-\varphi_{i}$. 
From (7) and (8) we have the expression for matrix Green's function 
$\hat G_{\omega}({\vec \rho}, {\bf v_F})$ as a function of the coordinate 
${\vec \rho}\in \Sigma$ ($\Sigma$ is the region of 2D rectangular weak link)
and the direction of ${\bf v_F}$. In fact we can write  

\begin{equation}
\hat G_{\omega}({\vec \rho},{\bf v}_F)= 
\hat G_{i\rightarrow j}, \ \   for \ \ {\bf v_F}\in \theta_{ij}({\vec \rho}),
\end{equation}
where we have introduced $\theta_{ij}({\vec \rho})$ as
the angle in which all $i\rightarrow j$ 
trajectories, passing through the point ${\vec \rho}$, are
confined (see Fig. 2, eq. (A1) in Appendix). The diagonal and off-diagonal 
terms of $\hat G_{i\rightarrow j}$ have the forms    

$$
g_{i\rightarrow j}=\frac{\omega}{\Omega}+\Delta_0 A_{i\rightarrow j}
$$
\begin{equation}
=\frac{\omega\cosh{(\omega t_{ji}+i\frac{\varphi_{ji}}{2})}
+\Omega\sinh{(\omega t_{ji}+i\frac{\varphi_{ji}}{2})}}{ \Omega
\cosh{(\omega t_{ji}+i\frac{\varphi_{ji}}{2})}+\omega\sinh{(\omega t_{ji}
+i\frac{\varphi_{ji}}{2})}},
\end{equation}
$$
f_{i\rightarrow j}=[\frac{\Delta_0}{\Omega}+(\Omega-\omega)
A_{i\rightarrow j}]\exp{[-2\omega(t-t_i)+i\varphi_i]}
$$
\begin{equation}
=\frac{\Delta_0 \exp{(\omega t_{ji}+i\frac{\varphi_{i}+\varphi_{j}}{2})}}
{\Omega\cosh{(\omega t_{ji}+i\frac{\varphi_{ji}}{2})}+\omega\sinh{(\omega 
t_{ji}+i\frac{\varphi_{ji}}{2})}}e^{-2\omega(t-t_i)}.
\end{equation}
In the limit $L,W<< \xi_0$ the expressions (10) and (11) for Green's functions 
are simplified and we have

\begin{equation}
g_{i\rightarrow j}=\frac{\omega\Omega+i\frac{1}{2}{\Delta_0}^2
\sin{(\varphi_{ji})}}
{ \omega^2+{\Delta_0}^2 \cos^2{(\frac{\varphi_{ji}}{2})}},
\end{equation}
\begin{equation}
f_{i\rightarrow j}=\frac{\Delta_0}{\Omega\cos{(\frac{\varphi_{ji}}{2})}
+i\omega\sin{(\frac{\varphi_{ji}}{2})}}\exp{(\frac{\varphi_{i}
+\varphi_{j}}{2})}.
\end{equation}
\par
We can obtain the retarded and advanced Green's functions, $\hat G^{R,A}
(\epsilon)$, by analytical continuation of Matsubara Green's 
function $\hat G(\omega)$ (eqs. (9)-(13)). The poles of diagonal component
of the retarded Green's function,
$g^{R}(\epsilon,{\vec \rho},{\bf v}_F)$, determine the energies of 
local Andreev states in the system. The local density of states in the 
normal layer is given by the furmula
\begin{equation}
{\cal N}(\epsilon,{\vec \rho})= N(0)
<Re\ g(\omega=-i\epsilon,{\vec \rho},{\bf v}_F)>.
\end{equation}
Using the expressions (9), (12) and the fact that,
${\theta_{ij}}({\vec \rho})={\theta_{ji}}({\vec \rho})$, in the
case of small junction, we obtain:

$$
{\cal N}(\epsilon,{\vec \rho},\{{\varphi}_i\})= 
N(0) \sum_{i\not=j} <Re\ g(\omega=-i\epsilon,{\vec \rho},{\bf v}_F)
>_{\theta_{ij}}
$$
$$
=N(0) \sum_{i\not=j}\theta_{ij}({\vec \rho})Re\ g_{i\rightarrow j}
(\omega=-i\epsilon)
$$
$$
=N(0) \sum_{i<j}\theta_{ij}({\vec \rho})Re [g_{i\rightarrow j}
(\omega=-i\epsilon)+
g_{j\rightarrow i}(\omega=-i\epsilon)]
$$
\begin{equation}
=\pi \Delta_0 N(0) \sum_{i<j}\theta_{ij}({\vec \rho})
\mid \sin{\frac{\varphi_{ji}}{2}}\mid 
\delta(\mid \epsilon\mid -\Delta_0 \cos{\frac{\varphi_{ji}}{2}}).
\end{equation}
We can also use eqs. (9) and (12) to obtain $<{\bf v}_F g>$ at a point of 
the $i$th side ${\vec \rho}_i$. Then, the resulting expression can be 
replaced in eq. (4)
to find the current density ${\bf j}({\vec \rho}_i)$. The calculation
of the current density at the arbitrary point of the normal rectangular 
will come in the next section and the appendix. Here we calculate 
the total current $I_i$ flowing into the $i$th bank.\\
Let us start with the case of a small junction ($L,W<<\xi_0$). 
In order to find $I_i$, we have to calculate the integral 
$I_i=\int_{(S_i)} {\bf j}({\vec \rho _{i}}).{\bf ds}_{i}$, 
where the integral is taken over the $i$th side of rectangular.\\
After calculation of $ {\bf j}({\vec \rho _{i}})$ from (A3) and (A5) and
taking the 
integral over ${\bf ds}_i$, we obtain for the current $I_i$

\begin{equation}
I_{i}= \frac{ep_F\Delta_0 d}{2\pi} \sum_{j=1} ^{4} \gamma_{ij}
\sin{(\frac{\varphi_{i}-\varphi_{j}}{2})}\tanh{[\frac{\Delta_0 
\cos{(\frac{\varphi_{i}-\varphi_{j}}{2})}}{2T}]}
\end{equation}
where $d=\sqrt{L^2+W^2}$ and $\gamma_{ij}=\gamma_{ji}$,  

$$
\gamma_{13}=1-\frac{k}{\sqrt{1+k^2}},
$$
$$
\gamma_{24}=1-\frac{1}{\sqrt{1+k^2}},
$$

\begin{equation}
\gamma_{12}=\gamma_{14}=\gamma_{23}=\gamma_{34}=
\frac{1}{2}(\frac{1+k}{\sqrt{1+k^2}}-1),
\end{equation}
are geometrical form factors which depend on the width to length 
ratio $k=W/L$. The positive sign of $I_i$ corresponds to the 
direction of the current from the normal layer to the $i$th bank. 
Note that $\sum_{i=1} ^{4} I_{i} =0$.\\
The formula (16) for current-phase relations generalizes the expression (1) 
to the case of a small mesoscopic 4-terminal junction. 
It follows from (17) that the form factor $\gamma_{ij}$ can not be 
factorized, i.e., presented in the form $\gamma_{ij}=\gamma_{i}\gamma_{j}$,
in contrast to the case of relation (1) where
$\gamma_{ij}=\frac{1}{R_{i}}.\frac{1}{R_{j}}$. This essential feature of the
current-phase relations reflects the nonlocal nature of the supercurrents in 
the mesoscopic multi-terminal Josephson junction.\par
The current-phase relations (16) are valid for 
arbitrary temperature $T$.
In the limiting cases of $T=0$ and temperature close to $T_c$ the expression 
(16) takes the forms

\begin{equation}
I_{i}= \frac{ep_F\Delta_0 d}{2} \sum_{j=1} ^{4} \gamma_{ij}
\sin{(\frac{\varphi_{i}-\varphi_{j}}{2})} \ \  for \ T=0,
\end{equation}
\begin{equation}
I_{i}= \frac{ep_F{\Delta_0}^2 d}{4\pi T_c} \sum_{j=1} ^{4} 
\gamma_{ij}\sin{(\varphi_{i}-\varphi_{j})} \ \ for \ T\simeq T_c.
\end{equation}
In the case of arbitrary lengths $L,W$ we restrict the consideration for 
the temperature close to $T_c$. In this case the current-phase relations 
similar 
to the expression (19) can be obtained. The difference is in geometrical 
form factors. In fact we have the result

\begin{equation}
I_{i}= \frac{ep_F{\Delta_0}^2 d}{4\pi T_c} 
\sum_{j=1} ^{4}{\tilde \gamma}_{ij}(k,L,W)\sin{(\varphi_{i}-\varphi_{j})}
\end{equation}
where the generalized form factors are given by

$$
{\tilde \gamma}_{41}= 
{\tilde \gamma}_{21}={\tilde \gamma}_{23}={\tilde \gamma}_{43}=
$$
$$
\frac{4}{\pi ^2\sqrt{1+k^2}}\int_{-\frac{k}{2}}^{\frac{k}{2}}
dy \int_{\arctan{(\frac{k}{2}+y)}}^{\frac{\pi}{2}}d\theta \cos{\theta} 
\sum_{n=0} ^{\infty}\frac{\exp{[-\frac{L(\frac{k}{2}+y)}
{\xi_N\cos{\theta}}(2n+1)]}}{(2n+1)^2},
$$

\begin{equation}
{\tilde \gamma}_{42}={\tilde \gamma}_{24}= \frac{4}{\pi ^2 \sqrt{1+k^2}}
\int_{-\frac{k}{2}}^{\frac{k}{2}}dy 
\int_{-\arctan{(\frac{k}{2}+y)}}^{ \arctan{(\frac{k}{2}-y)}}
d\theta \cos{\theta} \sum_{n=0} ^{\infty}\frac{\exp{[-\frac{L}
{\xi_N\cos{\theta}}(2n+1)]}}{(2n+1)^2},
\end{equation}
$$
{\tilde \gamma}_{12}={\tilde \gamma}_{14}={\tilde \gamma}_{32}
={\tilde \gamma}_{34}={\tilde \gamma}_{41}( L\rightarrow W, 
W\rightarrow L, k\rightarrow \frac{1}{k}),
$$
$$
{\tilde \gamma}_{13}={\tilde \gamma}_{31}=
{\tilde \gamma}_{42}( L\rightarrow W, 
W\rightarrow L, k\rightarrow \frac{1}{k}).
$$
Here $\xi_N=\frac{v_F}{2\pi T}$. In the limit $L,W<<\xi_0$, 
$\tilde \gamma_{ij}$ reduce to $\gamma_{ij}$.  

\section{Spatial distribution of supercurrents and induced order parameter}

In this section we will obtain the supercurrent density and
the induced order parameter
at an arbitrary point of the normal layer in the case of small
junction. At the given point of the normal layer ${\vec \rho}=x{\bf i}
+y{\bf j}$

$$
<{\bf v_F} g>= \sum_{i>j} (<{\bf v_F} g>_{\theta_{ij}} 
+<{\bf v_F} g>_{\theta_{ji}})
$$
\begin{equation}
=i\sum_{i>j} <{\bf v_F}>_{\theta_{ij}}
\frac{{\Delta_0}^2 \sin{\varphi_{ji}}}{{\omega}^2+{\Delta_0}^2
\cos^2{\frac{\varphi_{ji}}{2}}}
\end{equation}
where we have used 
$<{\bf v_F} g>_{\theta_{ij}}=
<{\bf v_F}>_{\theta_{ij}}g_{i\rightarrow j}, \ <{\bf v_F}>_{\theta_{ji}}
=-<{\bf v_F} >_{\theta_{ij}}$
and $ g_{j\rightarrow i}=g_{i\rightarrow j}^*$.\\
Replacing (22) in the eq. (4) the current density is obtained

\begin{equation}
{\bf j}({\vec \rho})=2\pi e N(0)\Delta_0 \sum_{i>j} <{\bf v_F} >_{\theta_{ij}}\sin{\frac{\varphi_{ji}}{2}}\tanh{(\frac{\Delta_0 \cos{\frac{\varphi_{ji}}{2}}}{2T})}
\end{equation}
The expression (23) describes the spatial distribution of the current 
density inside the normal layer. In order to find the explicit expression
for the coefficients $<{\bf v_F} >_{\theta_{ij}}$ in eq. (23), we have to 
consider 4 different regions in the normal 
rectangular and obtain ${\bf j}({\vec \rho})$ in each region 
separately (see Appendix). This calculation has been done in the 
appendix and the result for ${\bf j}({\vec \rho})$ is given by (A3) and
(A5). Here we write eq. (23) in the more transparent form. Let us 
introduce ${\hat \theta_{ij}({\vec \rho})}$ as the unit vector in
the direction of
the $i\rightarrow j$ trajectory passing through the bisector of
${\theta_{ij}}({\vec \rho})$; then, $<{\bf v}_F>_{\theta_{ij}} $ can
be wrirren as
\begin{equation}
<{\bf v}_F>_{\theta_{ij}}=\int_{0}^{{\theta_{ij}}}
\frac{d\theta}{2\pi}{\bf v}_F= \frac{v_F}{\pi}\sin{(\frac{\theta_{ij
}}{2})}
{\hat \theta_{ij}}.
\end{equation}
Combining eqs. (23) and (24), we obtain

\begin{equation}
{\bf j}({\vec \rho})=\frac{e p_F\Delta_0}{2\pi}
\sum_{i<j} \sin{(\frac{\theta_{ij}}{2})}
{\hat \theta_{ij}({\vec \rho})}\sin{(\frac{\varphi_{j}-
\varphi_{i}}{2})}
\tanh{[\frac{\Delta_0 \cos{(\frac{\varphi_{j}-\varphi_{i}
}{2})}}{2T}]}.
\end{equation}
In the way similar to what we have done for ${\bf j}({\vec \rho})$, 
the distribution of the induced order parameter can be obtained.
In this cas we need to calculate
the average of off-diagonal element of the matrix Green's function, 
$f_{\omega}({\bf v_F},{\vec \rho})$, on the direction of ${\bf v_F}$:

\begin{equation}
<f>= \sum_{i\not=j} <f>_{\theta_{ij}} 
=\sum_{i\not=j} \theta_{ij}({\vec \rho})f_{i\rightarrow j}.
\end{equation}
Replacing (26) in (3) and after the calculation, we obtain for 
$\Psi (x,y)=\Delta (x,y)/\lambda$:
 
\begin{equation}
\Psi (x,y) = \frac{\Delta_0}{\lambda}\sum_{i=1} ^{4} 
\theta_i(x,y) e^{i\varphi_i}.
\end{equation}
Here $ \theta_i(x,y)$ is the angle by which $i$th side is seen from
the point ${\vec \rho}\equiv (x,y)$. The angles
$\theta_i(x,y)$'s are given by the relations

$$
\theta_1=\pi-\alpha (x,y)-\alpha (-x,y), \ \theta_2=
\alpha (-x,-y)+\alpha (-x,y),
$$
\begin{equation}
\theta_3=
\pi-\alpha (x,-y)-\alpha (-x,-y), \ \theta_4=
\alpha (x,y)+\alpha (x,-y). 
\end{equation}
where the angle
\begin{equation}
\alpha (x,y)=\arctan{(\frac{k/2+y}{1/2+x})},
\end{equation}
is a function of the coordinate (normalized by $L$) and is shown in Fig. 6.\\
Eq.(27) expresses the fact that, inside the ballistic normal layer
region the linear superposition of four macroscopic wave functions 
(pair potentials) of the banks occurs,
where the weight of wave function of the $i$th bank is determinened
by the geometrical factor $\theta_i(x,y)$.

\section{Conclusions}

The present study considers a 4-terminal microstructure based on a
new class of mesoscopic Josephson junctions [9] which are fully phase
coherent and have comparable width and length. The microscopic theory of
the stationary coherent current states in ballistic multiterminals is 
developed.
\par
We have calculated the current-phase relations (CPR), i.e. the total currents
in each terminal as functions of the phases of the superconducting order 
parameter in all the banks.
These relations describe the behaviour of the system influenced
by the external transport currents or the applied magnetic fluxes. 
The essential difference between the CPR for mesoscopic
(expression (19)) and conventional (relation (1)) 4-terminals
consists in the structure of the coefficients
of coupling $\gamma_{ij}$.
In the mesoscopic case considered here these coefficients can not be 
factorized (presented in the form $\gamma_{ij}=\gamma_{i}\gamma_{j}$) 
for all indexes $i,j$ and arbitrary value of
the width to length ratio $k=W/L$. Here we only outline the new effect,
specific for the mesoscopic 4-terminal junction, which follows from such
nonlocal coupling of the currents. Let us consider the configuration shown in
Fig. 3. By using the CPR (19) with $\gamma_{ij}$ given by (17), 
it can be shown that an
applied magnetic flux through one of the rings produces magnetic flux in
the other ring even in the absence of an external flux through the 
other one. The
detailed theory of this effect will be reported in a separate publication.
\par
The physical properties of the interior of the mesoscopic 4-terminal
junction present the interest themselves. The  above calculated 
local density of
Andreev states, the current density and the order parameter distributions
depend on the phase differences between the four terminals and can be 
regulated
by the applied magnetic fluxes. In particular, for some values of the phases
$\varphi,\ \theta$ and $\chi$ (see Fig. 3) the "vortex state" inside the 
mesoscopic 2D weak link exists. Figures 4 and 5 present the plots
for distributions
of the absolute value of the induced order parameter 
and the supercurrent density in the case 
$\theta =\pi/2,\ \varphi =3\pi/2,\ \chi=0$.
The studying of the structure of induced order parameter and local density
of states, as well as the dynamical behaviour of the system will be the 
object of further investigation.\\

{\Large Acknowledgments}

The authers would like to acknowledge support for this research from
the Institute for Advanced Studies in Basic Scienses at Zanjan, IRAN.
We acknowledge to R.de Bruyn Ouboter and I.O.Kulik for usefull discussions.
\newpage

{\Large Appendix}\\

In this appendix we present expressions for the angles ${\theta}_{ij}$
and the vectors $<{\bf v}_F>_{{\theta}_{ij}}$. Using the expressions given here,
one can calculate the density of states
${\cal N}$ and the current density ${\bf j}$ 
[see eqs. (15) and (23)].\par

According to the classification of the trajectories in term of origin
and destination sides, there are $12$ different types of trajectories
which are $1\rightarrow 2, 1\rightarrow 3, 1\rightarrow 4, 2\rightarrow 3,
 2\rightarrow 4,
3\rightarrow 4$ and the corresponding reverse of these trajectories.
For a given point, depending on the position, some of these trajectories
do not take place. In this respect we can consider four different regions
in the normal rectangular: 
$I$ where $y<0, \mid y\mid>k\mid x\mid$,( $2\rightarrow 3, 
3\rightarrow 4$ and their reversed are absent) \\       
$II$ where $x\geq 0, \mid y\mid\leq kx$,($1\rightarrow 4, 
3\rightarrow 4$ and their reversed are absent), \\
$III$ where $y\geq 0, y>k\mid x\mid$,( $1\rightarrow 2, 
1\rightarrow 4$ and their reversed are absent) and \\ 
$IV$ where $x<0, \mid y\mid\leq kx$,( $1\rightarrow 2, 
2\rightarrow 3$ and their reversed are absent).\\
At the given point ${\vec \rho}$, for the absent trajectories we have
$\theta_{ij}=0$, and consequently the corresponding term
in the expressions of
${\cal N}$ and ${\bf j}$ [eqs. (15) and (23)] will vanish. 
We will calculate ${\bf j}$ in the given 
point of the region $II$ and then introduce the exchange rules of arguments
to obtain it in other regions.\\
Consider a point in region $II$; the possible (non-vanishing)
${\theta}_{ij}$ are drawn in Fig. 6 and, can be expressed in terms
${\theta}_{i}$'s (given by (28) and(29)) as 

$$
{\theta}_{12}=\frac{1}{2}({\theta}_{1}+{\theta}_{2}-{\theta}_{3}
-{\theta}_{4})
$$
$$
{\theta}_{13}=\frac{1}{2}({\theta}_{1}-{\theta}_{2}+{\theta}_{3}
+{\theta}_{4})
$$
$$
{\theta}_{23}=\frac{1}{2}(-{\theta}_{1}+{\theta}_{2}+{\theta}_{3}
-{\theta}_{4})
$$
$$
{\theta}_{24}={\theta}_{4}
\eqno{(A1)}
$$
Also we can use the relation, $<{\bf v}_F>_{{\theta}_{ij}}
=\int_{({\theta}_{ij})} \frac{d\theta}{2\pi}v_F(\cos{\theta}{\bf i}
+\sin{\theta}{\bf j})$, to obtain

$$
<{\bf v}_F>_{{\theta}_{12}}(x,y)=\frac{v_F}{2\pi}\{[\sin[\alpha (-x,-y)]
-\sin[\alpha (x,y)]]{\bf i}
$$
$$
+[\cos[\alpha (x,y)]-\cos[\alpha (-x,-y)]]{\bf j}\}
$$

$$
<{\bf v}_F>_{{\theta}_{13}}(x,y)=\frac{v_F}{2\pi}\{[\sin[\alpha (-x,y)]
-\sin[\alpha (-x,-y)]]{\bf i}
$$
$$
+[\cos[\alpha (-x,y)]+\cos[\alpha (-x,-y)]]{\bf j} \}
$$

$$
<{\bf v}_F>_{{\theta}_{23}}(x,y)=\frac{v_F}{2\pi}\{[\sin[\alpha (x,-y)]
-\sin[\alpha (-x,y)]]{\bf i}
$$
$$
+[\cos[\alpha (x,-y)]-\cos[\alpha (-x,y)]]{\bf j}\}
$$
$$
<{\bf v}_F>_{{\theta}_{24}}(x,y)=\frac{v_F}{2\pi}\{-[\sin[\alpha (x,y)]
+\sin[\alpha (x,-y)]]{\bf i}
$$
$$
+[\cos[\alpha (x,y)]-\cos[\alpha (x,-y)]]{\bf j}\}
\eqno{(A2)}
$$
with $\alpha (x,y)$ is given by eq. (29). The corresponding relations valid 
for other regions, can be obteined from (A1) and (A2), using
the appropriate rules 
of index and coordinate exchanging (see below).\\
Replacing eqs. (A1) and (A2) in (23), we obtain for current density in
a point of region
$II$
, ${\bf j}_{II}$

$$
{\bf j}_{II}(x,y) =
[-{\bf k}(x,y)+{\bf l}(x,y)]P_{13}
+[{\bf k}(-x,-y)-{\bf k}(x,y)]P_{12}
$$
$$
-[{\bf k}(x,y)+{\bf l}(-x,-y)]P_{24}
+[{\bf l}(-x,-y)-{\bf l}(x,y)]P_{23},
\eqno{(A3)}
$$
where
$$
{\bf k}(x,y)=\sin{\alpha}(x,y){\bf i}-
\cos{\alpha}(x,y){\bf j},
$$
$$
{\bf l}(x,y)=\sin{\alpha}(-x,y){\bf i}+
\cos{\alpha}(-x,y){\bf j},
\eqno{(A4)}
$$
and $P_{ij}=\frac{e p_F\Delta_0}{2\pi}
\sin{\frac{\varphi_{ji}}{2}}\tanh{(\frac{\Delta_0 
\cos{\frac{\varphi_{ji}}{2}}}{2T})}$. 
The current density in other regions is obtained from  ${\bf j}_{II}$ by
applying the following rules of phase and coordinate exchanging

$$
{\bf j}_{I}= {\bf j}_{II}[( x\rightarrow -y/k, y\rightarrow x/k, 
k\rightarrow 1/k);
({\bf i}\rightarrow -{\bf j}, {\bf j}\rightarrow {\bf i});
$$
$$
(\varphi_1\rightarrow \varphi_4, \varphi_2\rightarrow \varphi_1,
\varphi_3\rightarrow \varphi_2, \varphi_4\rightarrow \varphi_3)], 
$$
$$
{\bf j}_{III}= {\bf j}_{II}[( x\rightarrow y/k, y\rightarrow -x/k, 
k\rightarrow 1/k);
({\bf i}\rightarrow {\bf j}, {\bf j}\rightarrow -{\bf i});
$$
$$
(\varphi_1\rightarrow \varphi_2, \varphi_2\rightarrow \varphi_3,
\varphi_3\rightarrow \varphi_4, \varphi_4\rightarrow \varphi_1)],
\eqno{(A5)}
$$
$$
{\bf j}_{IV}= {\bf j}_{II}[( x\rightarrow -x, y\rightarrow -y, 
k\rightarrow k);
({\bf i}\rightarrow -{\bf i}, {\bf j}\rightarrow -{\bf j});
$$
$$
(\varphi_1\rightarrow \varphi_3, \varphi_2\rightarrow \varphi_4,
\varphi_3\rightarrow \varphi_1, \varphi_4\rightarrow \varphi_2)].
$$
The same relations as (A5) can be used for 
${\theta}_{ij}$
and $<{\bf v}_F>_{{\theta}_{ij}}$ (the phase exchanges have to be
replaced
by corresponding index exchanges).

\newpage
{\large Figure Captions}

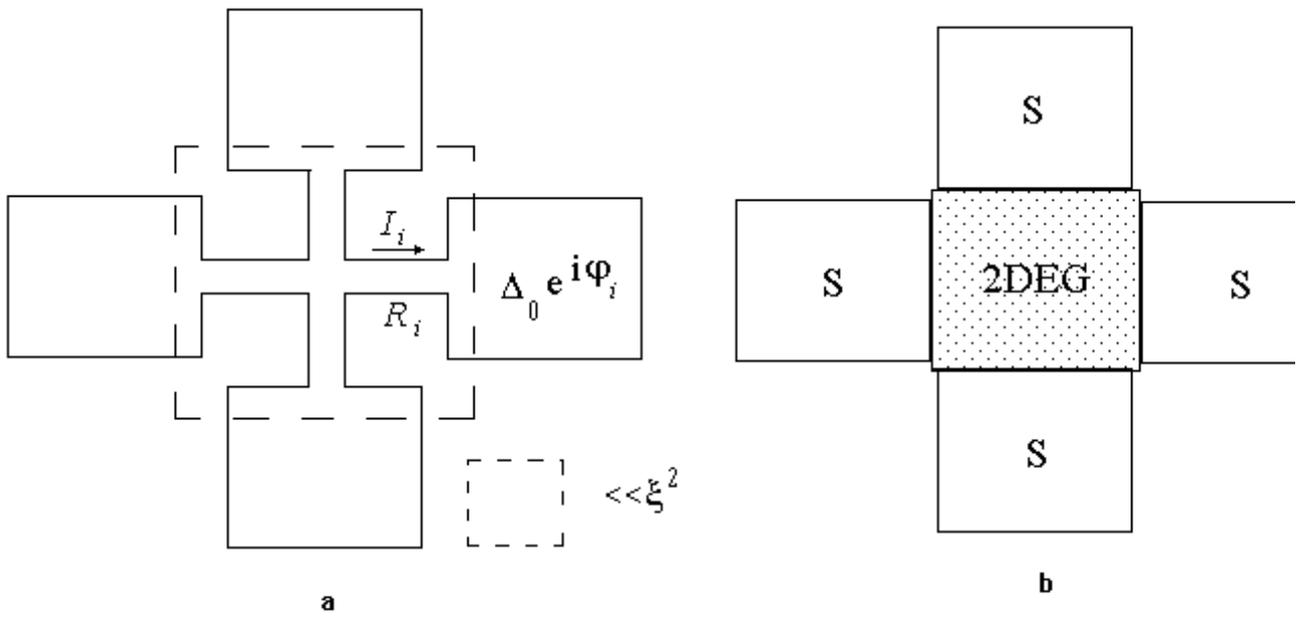
Figure 1. (a) The superconducting 4-terminal Josephson junction.
Four coupled superconductind microbridges, going from a week point
to the massive superconducting banks ($R_i$ is the normal resistance 
of the $i$th filament and $\xi (T)$ is the coherence length).\\
(b) The mesoscopic 4-terminal Josephson junction. Four bulk superconductors
are weakly coupled through a rectangular of two-dimensional electron gas 
(2DEG). 
\par
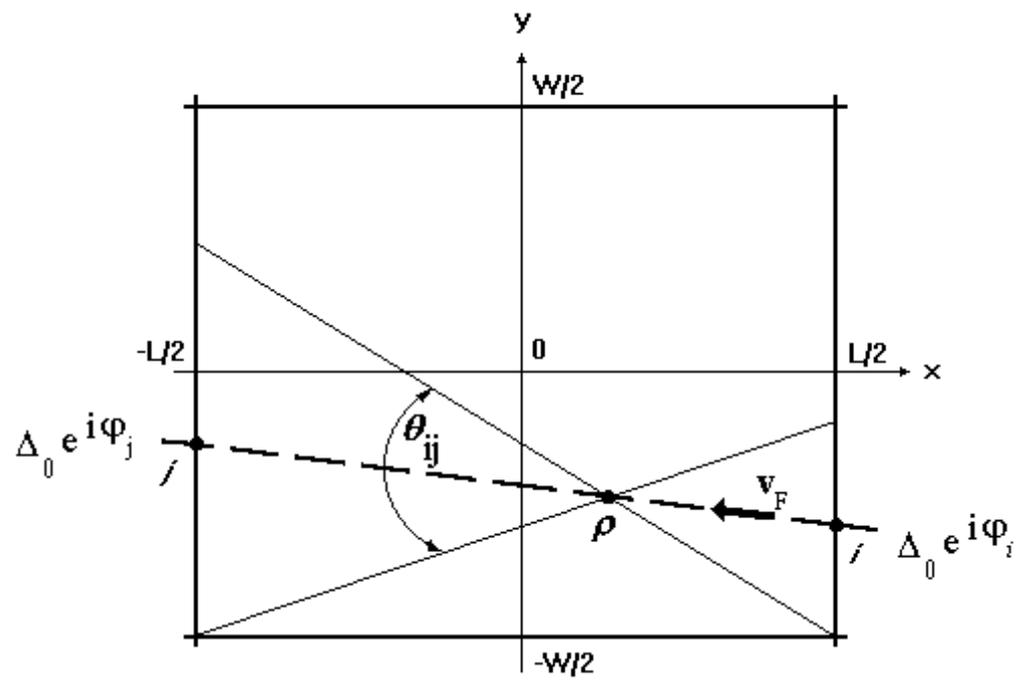
Figure 2. Dashed line is $i\rightarrow j$ trajectory passing through the
point ${\vec \rho}$. All of this type trajectories are confined in the angle
${\theta_{ij}}$. $L$, $W$ are length and width of the rectangular.
\par
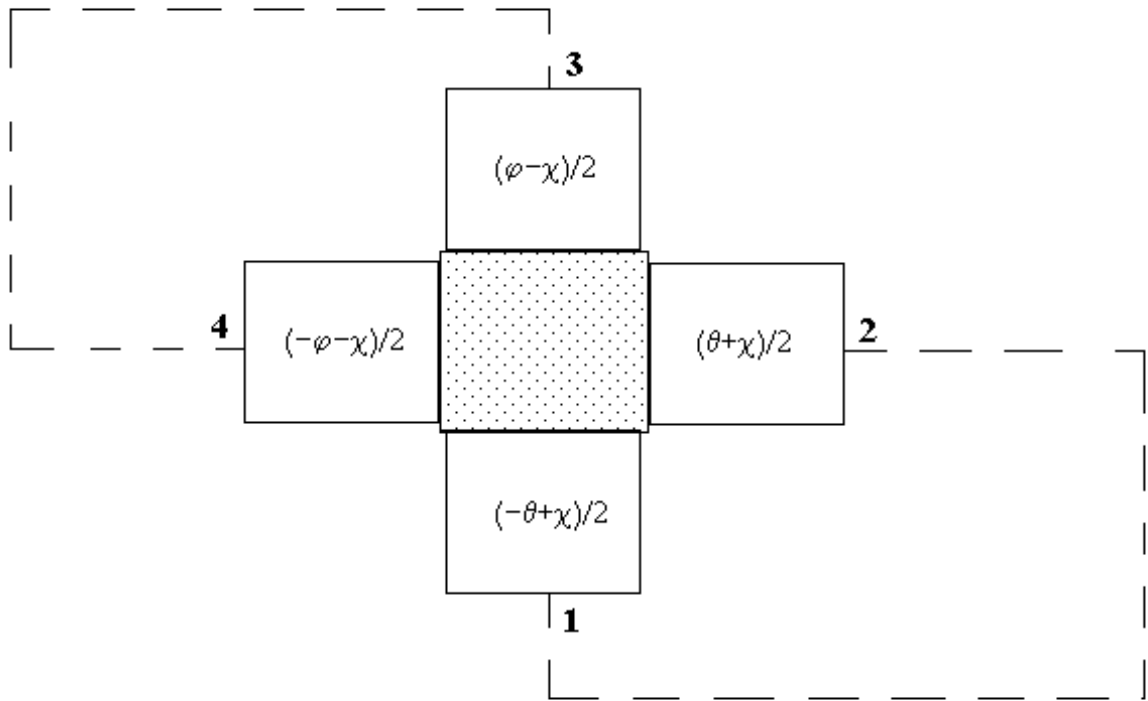
Figure 3. A configuration of the mesoscopic 4-terminal Josephson junction.
The terminals 1 with 2, and 3 with 4 are short-circuited by
the superconducting 
rings (dashed lines). The phase differences are $\theta=\varphi_2
-\varphi_1,\  \varphi=\varphi_3-\varphi_4,\  \chi=
\frac{\varphi_1+\varphi_2}{2}-\frac{\varphi_3+\varphi_4}{2}$.
\par
Figure 4. Absolute value of the induced order parameter $\mid \psi(x,y)\mid$
is plotted verticaly for the values of phase differences
$\theta=\varphi_2
-\varphi_1=\pi/2,\  \varphi=\varphi_3-\varphi_4=3\pi/2,\  \chi=
\frac{\varphi_1+\varphi_2}{2}-\frac{\varphi_3+\varphi_4}{2}=0$.
The lines of $\mid \psi(x,y)\mid=const.$ are shown.
\par
Figure 5. Vector field plot of the supercurrent density, ${\bf j}(x,y)$,
inside the normal layer. The valuese of phase differences are the same as
the Fig. 4.
\par
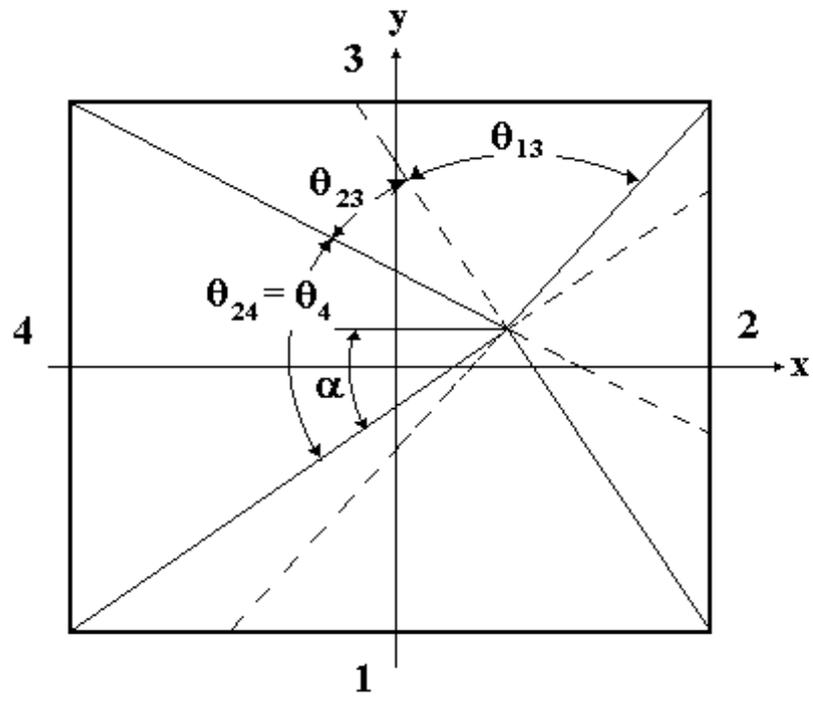
Figure 6. The angles ${\theta_{ij}}$ for a point in the region $II$.
We have just shown ${\theta_{13}}$, ${\theta_{23}}$, ${\theta_{24}}=
{\theta_{4}}$ and also the angle $\alpha$.

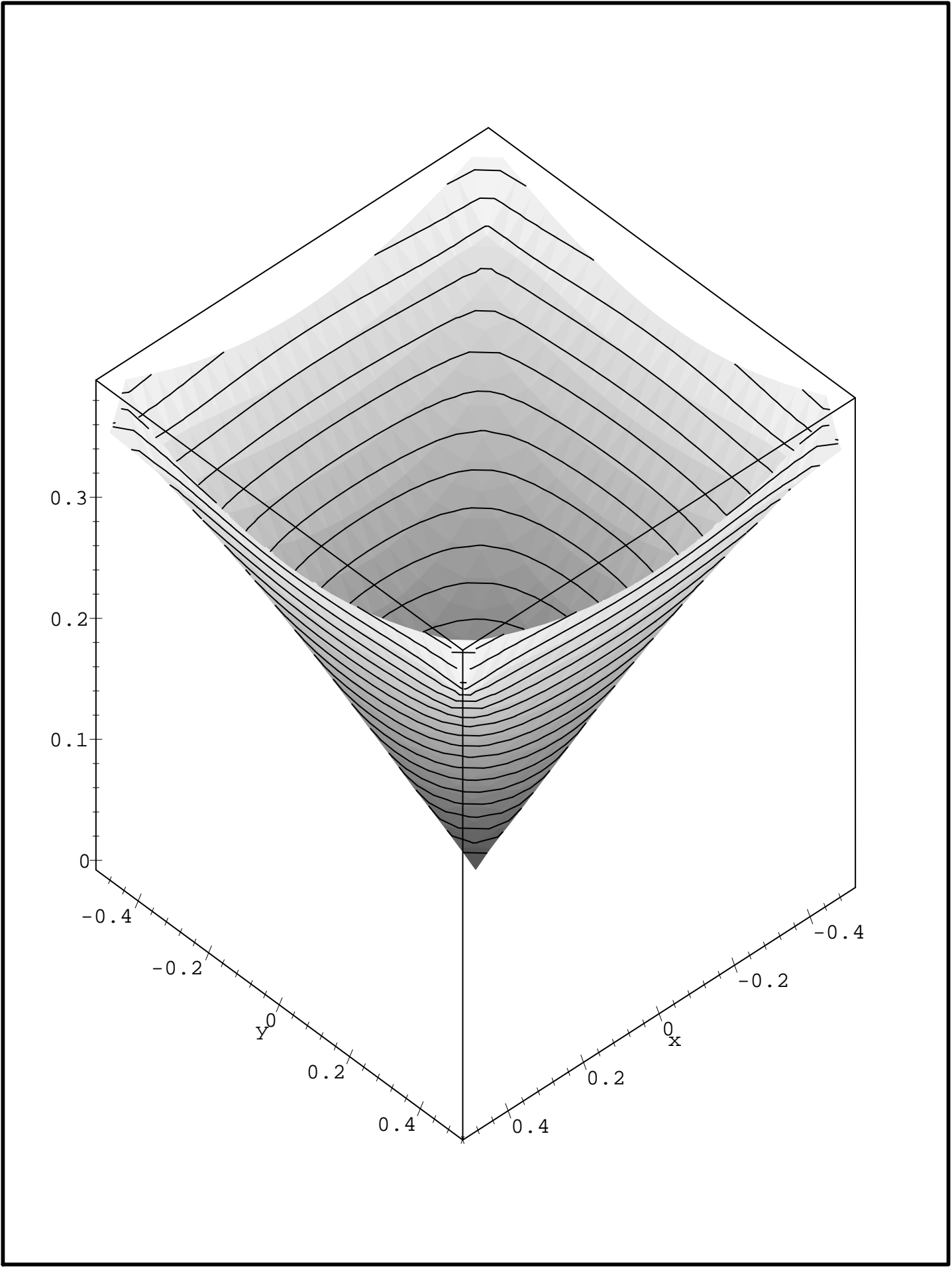
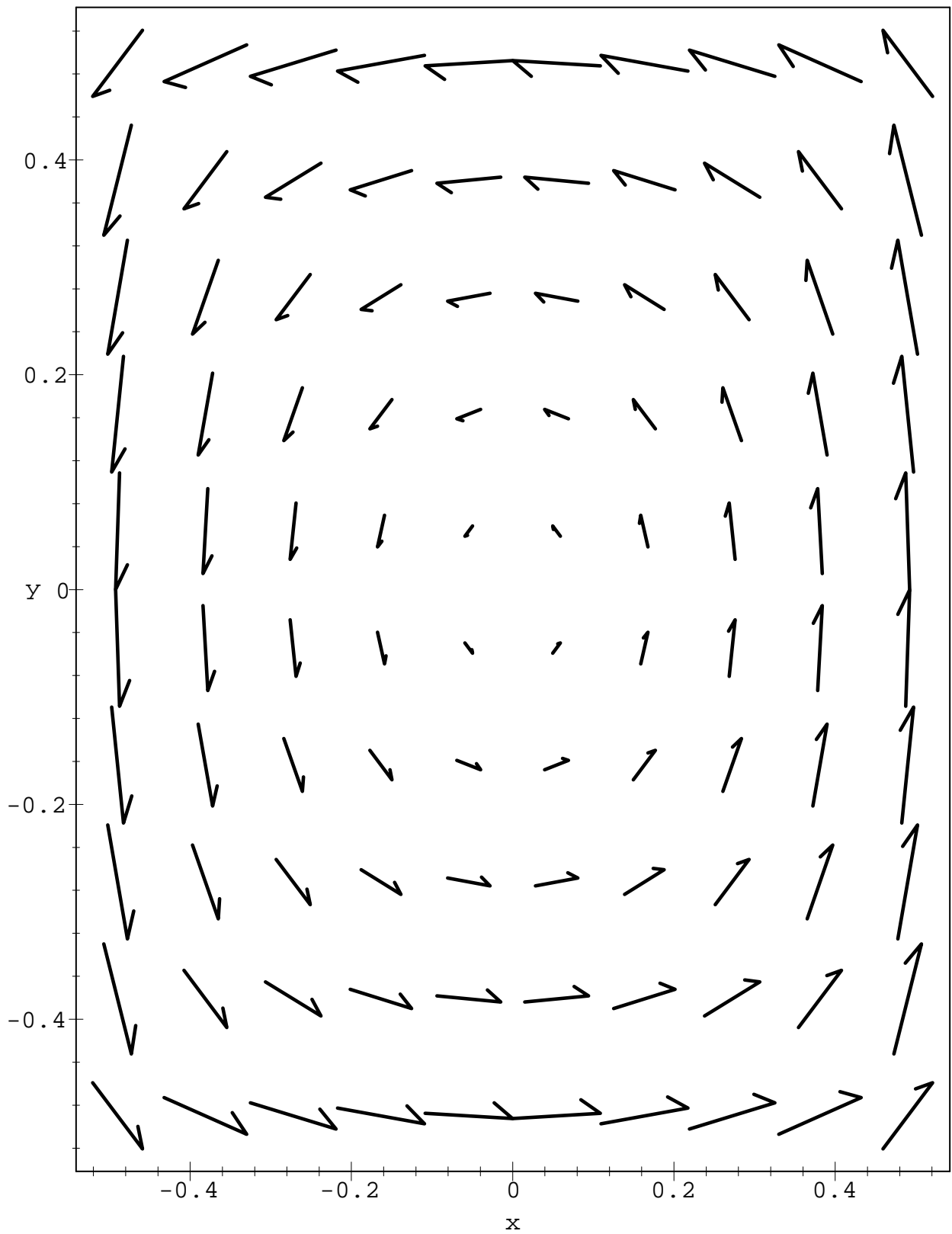
\end{document}